# The Protons in Primary Cosmic Rays in the Energy Range $10^{15}$-$10^{17}$ eV According to Data from the PAMIR Experiment

V.S. Puchkov[1], E.A. Kanevskaya[1], V.M. Maximenko[1], R.A. Mukhamedshin[2], S.E. Pyatovsky[1], and others

[1]*Lebedev Physical Institute, Russian Academy of Sciences*

[2]*Institute for Nuclear Research, Russian Academy of Sciences*

vgsep@ya.ru

**Abstract:** Adjusted data on the fraction of protons in the mass composition of primary cosmic rays (PCRs) in the energy range of $10^{15}$-$10^{17}$ eV are presented. Adjustments are made according to detailed calculations of the response of the X-ray emulsion chamber in the PAMIR experiment. It is demonstrated that the fraction of protons in a PCR is 16-18% for $E_0 \approx 10^{15}$-$10^{16}$ eV and does not change within the error for $E_0 \approx 10^{16}$-$10^{17}$ eV.

**Keywords:** cosmic rays, mass composition, PAMIR experiment.

**1 Introduction**

The mass composition of PCRs in the energy range of $10^{15}$-$10^{17}$ eV remains an unsolved problem. Analysis of the data from most experiments to investigate extensive air showers (EASes) leads to conclusions as to the rapid growth of the fraction of intermediate and heavy nuclei in PCRs in the energy range of $10^{15}$-$10^{16}$ eV [1]. Data from experiments with X-ray emulsion chambers sensitive to the fraction of protons in a PCR testifies to the presence of a large fraction of protons in the mass composition of PCRs over the range of $10^{15}$-$10^{17}$ eV. At lower energies ($E \approx 10^{14}$ eV), the fraction of protons in PCRs is ~ 30 %, according to data from balloon experiments; this does not contradict the data on EAS and X-ray emulsion chambers.

Detailed calculations for the response of X-ray emulsion chambers upon the registration of gamma-families have been performed in the recent years as part of the PAMIR experiment, using the GEANT 3.21 code in order to solve this contradiction; the code has been adapted for calculating the measurement procedures used in the PAMIR experiment. The influence of subthreshold cascades, the mutual influence of close fogging spots, and the possibility of cascade discrimination against a heightened background are all taken into account in the calculations.

Detailed analysis of the response of an X-ray emulsion chamber to the registration of different characteristics of gamma-families sensitive to the fraction of protons in PCR [2,3] allowed us to adjust the estimate of the fraction of protons in PCR in the energy range of 1-100 PeV.

## 2 Experimental

The gamma-families with $\Sigma E_\gamma \geq 100$ TeV registered in the PAMIR experiment in a thin lead X-ray emulsion chamber (6 cm of lead) were used for our analysis. Densitometry measurements were performed mainly at a depth of 5 cm of lead, which corresponds to the depth of the development of electromagnetic cascades 9-11 c.u. (with allowance for the family angle of incidence). Gamma-families were selected according to the following criteria: $\Sigma E_\gamma \geq 100$ TeV, $E_\gamma \geq 4$ TeV, $n_\gamma \geq 3$, $R \leq 15$ cm. We used 1003 events with total exposure $ST = 2635$ m$^2$ to calculate the intensity of gamma-families. The 61 gamma-families with halos of $\Sigma E_\gamma \geq 500$ TeV registered in the X-ray emulsion chamber of the PAMIR experiment over an exposure time of $ST = 3000$ m$^{-2}$ year$^{-1}$ were also used in our analysis. These were events in which close gamma-cascades in the central part of families overlap, forming a large diffusion fogging spot (a halo) with high optical density. Single-center and multi-center halos were observed in experiments. Calculations showed that single-center halos are formed mainly by protons. Halos formed by heavy nuclei are, as a rule, multi-center. The criteria for selecting gamma-families with halos are as follow:

(1) $\Sigma E_\gamma \geq 500$ TeV;

(2) the area of halo bounded by isodensity line $D = 0.5$, $S_{D = 0.5} \geq 4$ mm$^2$;

(3) in the case of a multi-center halo, $\Sigma S_{iD = 0.5} \geq 4$ mm$^2$ when $S_{iD = 0.5} \geq 1$ mm$^2$.

Calculations show that gamma-families with $\Sigma E_\gamma = 100\text{-}400$ TeV are formed mainly by primary particles with an energy of $10^{15}\text{-}10^{16}$ eV, while events having halos with $\Sigma E_\gamma \geq 500$ TeV are formed by primary particles with an energy of $E_0 \geq 10^{16}$ eV.

## 3 Model calculations

The MC0 model [4] developed for the PAMIR experiment in the range of primary energies $2 \times 10^{14}\text{-}3 \times 10^{18}$ eV was used to simulate gamma-families in the atmosphere. Calculations were performed using the primary spectrum obtained in the KASCADE and Tibet [5] experiments over an angular range of 0-45°. The mass composition in the MC0 model is given in Table 1.

Table 1. Fraction of light nuclei in PCR mass composition in the MC0 model.

| $E$, eV | $10^{15}$ | $10^{16}$ | $10^{17}$ |
|---|---|---|---|
| $p$, % | 33 | 26 | 20 |
| $\alpha$, % | 22 | 17 | 15 |

## 4 Intensity of gamma-families

Calculations with different models (MC0, QGSJET) demonstrate that experiments with X-ray emulsion chambers are sensitive mainly to the fraction of protons in PCRs. Around 80% of all events are due to primary protons, while helium nuclei create 10% of all gamma-families. Primary nuclei heavier than helium contribute no more than 10% to the total. This conclusion is virtually independent of the applied models, so a comparison of the intensities of gamma-families in the experiment and calculations is a good independent estimate of the fraction of protons in PCRs in the range of 1-100 PeV. To prove the above, we consider the spatial distribution of gamma-cascades in the families.

According to calculations, the average size of family $R$ formed by a proton is indeed 1.8 cm; for events formed by iron nuclei, it is 4.5 cm. Experimental and calculated R distributions are shown in the figure. Gamma-families with energy $\Sigma E_\gamma$ = 100-400 TeV in which reliable identification of all cascades is possible were used to construct the distributions. At high family energies, a halo can occur in the central part of an event; low-energy gamma-cascades can be lost in these halos, and determining $R$ can be complicated. It can be seen from the figure that the experimental $R$ distribution is well described by the set of spatial distributions of events formed by primary protons. The average size of a family in the PAMIR experiment is $R_{\text{Pamir}}$ = (2.28 ± 0.06) cm, which is somewhat higher than the data of the MC0 model, $R_{\text{MC0}}$ = (2.01 ± 0.03) cm, and virtually coincides with the average size of events formed by helium nuclei, $R_{\text{He}}$ = (2.37 ± 0.09) cm. However, an analysis of the response of X-ray emulsion chambers shows that the spatial distribution of cascades in a family as it passes through a chamber increases by 15 %. The overestimation of $R$ in the experiment is associated with the registration, due to fluctuations, of a large number of families with $\Sigma E_\gamma$ close to the threshold energy ($\Sigma E_{\gamma\text{thr}}$ = 100 TeV) that have broader spatial distributions. The true size of the gamma-families in the experiment is $R_{\text{Pamir}}$ = (1.94 ± 0.06) cm, which proves the conclusion made earlier as to the leading role of primary protons in the formation of gamma-families registered in X-ray emulsion chambers.

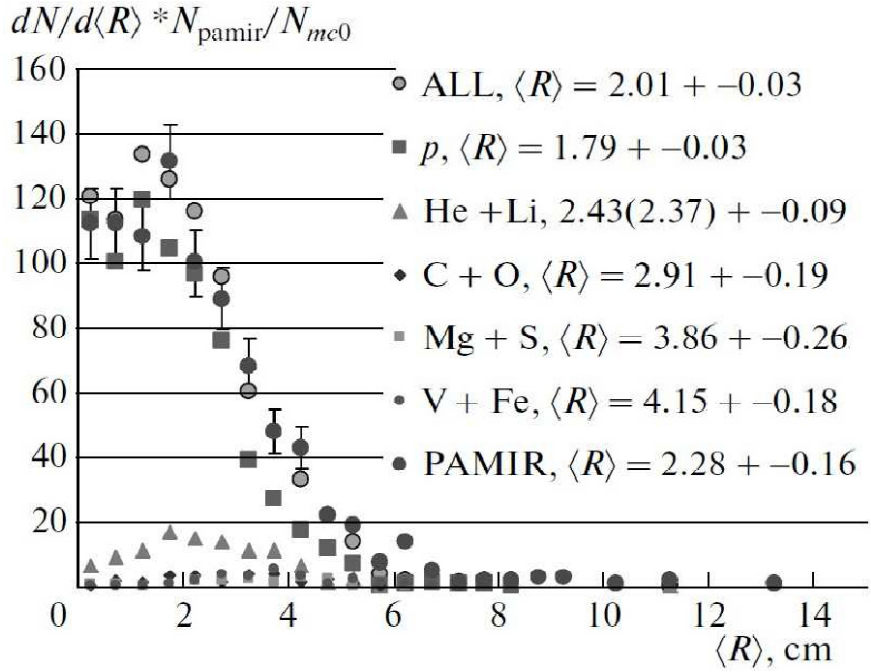

Figure. Distribution $dN_f/d\langle R_f \rangle$. Distributions for MC0 are normalized to the number of families in the PAMIR experiment.

The vertical intensity of gamma-families in the PAMIR experiment was $I_{vert}$ ($\Sigma E_\gamma \geq 100$ TeV, 600 g cm$^{-2}$, 0) = 0.62 ± 0.04 m$^{-2}$ year$^{-1}$ sr$^{-1}$ [6]. However, calculations of the X-ray chamber response showed that the experimentally observed event intensity was overestimated. The ratio of the number of gamma-families striking the X-ray chamber to the registered number of families is 63 %. The intensity of gamma-families in the PAMIR experiment is thus $I_{vert}$ ($\Sigma E_\gamma \geq 100$ TeV, 600 g cm$^{-2}$, 0) = 0.39 ± 0.03 m$^{-2}$ year$^{-1}$ sr$^{-1}$.

To match the experimental data and calculations, the fraction of protons in the model must be reduced from 33 % to (18 ± 2) % if excessive protons are replaced by nuclei of the iron group. If the excessive protons are replaced by helium nuclei, the fraction of protons in a PCR falls to (16 ± 2) %, and the fraction of helium rises to 34 %. The fraction of protons in the mass composition of PCRs in the energy range of $10^{15}$-$10^{16}$ eV is thus 16-18 % ± 2 %.

The intensity of gamma-families with halos $\Sigma E_\gamma \geq 500$ TeV obtained in the experiment is compared to the intensity of gamma-families with halos $\Sigma E_\gamma \geq 400$ TeV obtained in calculations using the MC0 model to estimate the fraction of protons in a PCR in the range of $E_0 \geq 10^{15}$-$10^{16}$ eV. The overestimation of the energy of gamma-families with $\Sigma E_\gamma \approx 500$ TeV following from our analysis of X-ray chamber response is thus taken into account.

The calculated value of global intensity $I_{glob}$ in the model is 75 events, while the experimental value is just 61 events. The experimental data thus agree with our calculations if the fraction of protons in a PCR for $E_0 \geq 10^{16}$ eV lies in the range of (16-18) ± 3 %. This range depends on to which group of nuclei the excess of primary protons assumed in the MC0 model is transferred.

For verify the correctness of the proposition that at high energies ($E_0 \geq 10^{16}$ eV) most events with halos are formed by primary protons, we compare the fraction of multi-center halos in the experiment and calculations. Table 2 gives the fraction of multi-center halos among gamma-families formed by different groups of nuclei.

Table 2. Fraction of multi-center halos in gamma-families formed by different nuclei.

| P | α | C | Fe | PAMIR |
|---|---|---|---|---|
| 0.25 ± 0.03 | 0.45 ± 0.09 | 0.59 ± 0.11 | 0.70 ± 0.12 | 0.23 ± 0.07 |

It can be seen from the table that gamma-families with halos are almost entirely generated by primary protons with the possibility of additions from helium nuclei.

## 5 Conclusions

Detailed analysis of X-ray emulsion chamber response shows that the intensity of gamma-families in the PAMIR experiment was overestimated. The true intensity is just 63 % of the one observed, resulting in the need to adjust the estimate of the fraction of protons in the mass composition of PCRs at $E_0 \approx 10^{15}$ eV. In this case, the intensity of families with halo remains virtually the same. According to the data of the PAMIR experiment, the range of values of the fraction of protons is 14-20 % and remains virtually the same up to $E_0 \approx 10^{17}$ eV. In the energy range of $10^{16}$-$10^{17}$ eV, the admissible fraction of protons in a PCR is 13-21 %.